\documentclass{emulateapj}

\shorttitle{Quasar Luminosity Evolution}
\shortauthors{Singal et al.}

\slugcomment{Accepted to ApJ}

\begin{document}

\title{ON THE RADIO AND OPTICAL LUMINOSITY EVOLUTION OF QUASARS}

\author{J. Singal\altaffilmark{1}, V. Petrosian\altaffilmark{1}$^,$\altaffilmark{2}, A. Lawrence\altaffilmark{3}, {\L}. Stawarz\altaffilmark{4}$^,$\altaffilmark{5}}

\altaffiltext{1}{Kavli Institute for Particle Astrophysics and Cosmology\\SLAC National Accelerator Laboratory and Stanford University\\382 Via Pueblo Mall, Stanford, CA 94305-4060}
\altaffiltext{2}{Also Departments of Physics and Applied Physics}
\altaffiltext{3}{University of Edinburgh Institute for Astronomy\\Scottish Universities Physics Alliance (SUPA)\\Royal Observatory, Blackford Hill, Edinburgh UK}
\altaffiltext{4}{Institute of Space and Astronautical Science (ISAS)\\ Japan Aerospace Exploration Agency (JAXA),\\ 
3-1-1 Yoshinodai, Chuo-ku, Sagamihara, Kanagawa 252-5510 Japan}
\altaffiltext{5}{Astronomical observatory of the Jagiellonian University\\ ul. Orla 171, 30-244 Krak\'ow, Poland}

\email{jsingal@stanford.edu}

\begin{abstract}
We calculate simultaneously the radio and optical luminosity evolutions of quasars, and the distribution in radio loudness $R$ defined as the ratio of radio and optical luminosities, using a flux limited data set containing 636 quasars with radio and optical fluxes from White et al.  We first note that when dealing with multivariate data it is imperative to first determine the true correlations among the variables, not those introduced by the observational selection effects, before obtaining the individual distributions of the variables.  We use the methods developed by Efron and Petrosian which are designed to obtain unbiased correlations, distributions, and evolution with redshift from a data set truncated due to observational biases.  It is found that the population of quasars exhibits strong positive correlation between the radio and optical luminosities.  With this correlation, whether intrinsic or observationally induced accounted for, we find that there is a strong luminosity evolution with redshift in both wavebands, with significantly higher radio than optical evolution.  We conclude that the luminosity evolution obtained by arbitrarily separating the sources into radio loud ($R>10$) and radio quiet ($R<10$) populations introduces significant biases that skew the result considerably.  We also construct the local radio and optical luminosity functions and the density evolution.  Finally, we consider the distribution of the radio loudness parameter $R$ obtained from careful treatment of the selection effects and luminosity evolutions with that obtained from the raw data without such considerations.  We find a significant difference between the two distributions and no clear sign of bi-modality in the true distribution for the range of $R$ values considered.  Our results indicate therefore, somewhat surprisingly, that there is no critical switch in the efficiency of the production of disk outflows/jets between very radio quiet and very radio loud quasars, but rather a smooth transition.  Also, this efficiency seems higher for the high-redshift and more luminous sources in the considered sample.

\end{abstract}


\section{Introduction} \label{intro}

The optical emission of quasars, or active galactic nuclei (AGN) is dominated by the radiation of the plasma accreting onto supermassive black holes, while the radio emission is dominated by the plasma outflowing from the black hole/accretion disk systems.  Hence different but complementary information can be gathered in both photon energy ranges regarding the cosmological evolution of AGNs and its relation to structure formation in the Universe.  It is therefore important to analyze in detail redshift distributions of quasars in both frequency regimes, investigating carefully any possible differences between these two.

The rapid evolution of active galaxies identified in radio catalogs as `quasi-stellar radio sources' (or QSRs) in the redshift range $z \lesssim 2$ was established soon after their discovery \citep[e.g.][]{Schmidt68}. Subsequent optical discoveries of similar sources, most of which had no detectable radio emission, lead to the emergence of the class of `radio quiet quasars' (or `quasi-stellar objects'; QSOs for short e.g. \citet{Osterbrock89}). These optically-selected sources also showed similar strong evolutionary trends, similar to the radio-selected ones.  These evolutions are modeled as density evolution, luminosity evolution or a combination of the two in numerous works \citep[e.g.][]{Schmidt68,P73,Marshall83,DP90,MP99,Willott01} and can be designated as the evolution of the luminosity function (LF, for short).  

By now the evolution of the LF has been described not only for optical and radio luminosities  but also for X-ray, infrared, and bolometric luminosities \citep[e.g.][]{Ueda03,Richards06,Matute06,Hopkins07,Croom09}.  Most of these studies have treated the evolution with a bi-variate function  $\Psi_i (L_i,z)$, where $L_i$ is the luminosity (or, in this case, luminosity spectral density) in some photon energy range, e.g. $L_i=L_{\rm opt}$ or $ L_{\rm rad}$.  The shape of the LF and its evolution are usually obtained from a flux limited sample $f_i > f_{m,i}$ with $L_i = 4 \, \pi \, d_L^2(z) \, (1/K_i(z)) \, f_i$, where $d_L$ is the luminosity distance and $K_i(z)$ stands for the K-correction.  For a power law emission spectrum of index $\varepsilon_i$ defined as $f_i \propto \nu^{-\varepsilon_i}$, one has $K_i(z) = (1+z)^{1-\varepsilon_i}$.  

However, because no matter how a quasar is discovered, optical observations are required for determination of redshift, then the flux limit of optical observations ($f_{m,opt}$) and the optical luminosity enter the picture, so that one now must consider the joint  LF and its evolution  $\Psi (L_{\rm opt},  L_i, z)$, a tri-variate function with $L_i = L_{\rm rad}$ or $L_{X}$, for example. 

In general, the first step required for investigation of a multivariate distribution is the determination of whether the variables of the distributions are correlated or are statistically independent.  For example, in the case of a single LF the correlation between $L$ and $z$ is what we call luminosity evolution, and independence of these variables would imply absence of such evolution.  Mathematically, independence means that the function is separable $\Psi_i(L_i,z)=\psi_i(L_i) \times \rho(z)$, in which case one is left with the determination of a single variable LF $\psi_i(L_i)$ and the density evolution $\rho(z)$.  As shown by \citet{P92} the most exact nonparametric method for this task from a flux limited (or a more generally truncated) sample is the \citet{L-B71} method.  However, this simple and elegant method cannot be used for cases when variables are correlated (e.g. when there is luminosity evolution).  \citet{EP92,EP99} (EP for short) developed  new methods for determination of the existence of correlation or independence of the variables from a flux limited and more generally truncated data set, and prescribed how to remove the correlation by defining new and independent variables [say $L'_i \equiv L_i/g_i(z)$ and $z$, where the function $g_i(z)$ describes the luminosity evolution] and then how to determine the mono-variate functions  $\psi_i(L'_i)$ and $\rho(z)$.  Thus, one can determine both the luminosity and density evolutions $g_i(z)$ and $\rho(z)$, as well as the LF at any redshift.\footnote{It should be noted that here we assume that the shape of the LF is constant; e.g. power law indices describing the LF are independent of $z$. In general, shape variations can affect the test of independence.  For a sufficiently large  sample the importance of these effects can be determined and accounted for. This is beyond the scope of this paper.}  

In the case of quasars with the optical and some other band luminosity, we have at least a tri-variate function.  In this case one must determine not only the correlations between the redshift and individual luminosities (i.e. the two luminosity evolutions) but also the possible correlation between the two luminosities, before individual distributions can be determined.  Knowledge of these correlations and distributions are essential for not only constraining robustly the cosmological evolution of active galaxies, but also for interpretation of related observations, such as the extragalactic background radiation \citep[e.g.][]{Singal10,Hopkins10}.

Another related aspect of this subject, which has attracted considerable attention over the years, is the distribution of the`radio-loudness parameter' for the quasar population, and the distinction between so-called `radio loud' (RL for short ) and `radio quiet' (RQ for short) quasars.   The question of whether there are two distinct populations was addressed soon after the discovery of quasars using small samples with radio flux limits greater than one Jy. Initially it was found that the distribution of the ratio of radio to optical luminosity $R \equiv L_{\rm rad}/L_{\rm opt}$, the so called ``radio loudness parameter'',  was a fairly broad power law with index $\beta_R\sim -2.3$  in the range $2.8<{\rm \log}R<5.2$ \citep{Schmidt72,P73}.  At that time this ratio was defined for the radio luminosity at $\nu_{\rm rad}=0.5$ GHz and optical luminosity at 2500 \AA \, (or the frequency $\nu_{\rm opt}=1.2 \times 10^{15}$ Hz).  Nowadays it is defined with radio luminosity at 5 GHz so that for a mean radio spectral index $\varepsilon_{\rm rad}\sim 0.6$ the old data would be in the range 2.2 to 4.6 of the modern definition of $\log R$.\footnote{The fiducial cosmological model used at that time, namely the Einstein-De Sitter model, was also different than the currently accelerating models.  However, this will affect the values of the individual luminosities but not the ratio $R$.  We also note that some other authors use B-band optical fluxes in defining the radio loudness parameter, but the difference is not large, resulting in a change in the ratio by a factor of 1.33, for the assumed optical spectral index $\varepsilon_{\rm opt}=-0.5$.}  Later, however, the survey limits were extended to much lower fluxes (specially in the radio domain), and this has resulted in a much wider range of the ratio that extends to values well below one, namely $-3< {\rm \log} R<5$.  Within this broader range, weak hints of the bi-modality described by \citep{Kellerman89} suggested that $\log R = 1$ could be chosen as the radio loud/quiet demarcation value.  Using this value for the division between RL and RQ quasars, the differences between the two classes have been investigated, including the possibility of distinct cosmological evolution of the RL and RL populations \citep[eg][]{Miller90,Goldschmidt99,Jiang07}.   Still, the more recent analyses of different samples of objects reported in the literature so far gave rather inconclusive results on whether any bi-modiality in the distribution of the radio loudness parameter for quasars is inherent in the population  \citep[see][]{Ivezic02,Cira03,Ivezic04}

There have been many papers dealing with this ratio and RL vs RQ issue, as well as luminosity ratios at other wavelengths, e.g. IR/radio, Optical/X-ray etc.  However, none of these works have dealt with the intrinsic distribution (and/or evolution) of the ratio, which is related to the tri-variate LF $\Psi (L_{\rm opt},  L_{\rm rad}, z)$ by\footnote{Equation \ref{Gtrue} arises because by definition $\int {G_R(R,z) \, dR} = \int \int {\Psi(L_{\rm opt}, \, L_{\rm rad} \, z) \, dL_{\rm opt} \, dL_{\rm rad}}$, and from the definition of $R$, $dL_{\rm rad} = L_{\rm opt} \, dR$ and $dL_{\rm opt} = -(L_{\rm rad} / R^2) \, dR$.}
\begin{eqnarray}
G_R(R,z) = \int_0^{\infty} { \Psi (L_{\rm opt},  R \,\, L_{\rm opt}, z) \, L_{\rm opt} \, dL_{\rm opt} } \nonumber \\ = \int_0^{\infty} { \Psi \left( {L_{\rm rad} \over R},  L_{\rm rad}, z \right) \, L_{\rm rad} \, {{ dL_{\rm rad} } \over {R^2}} } 
\label{Gtrue}
\end{eqnarray}
These works did not take the observational selection effects properly into consideration, nor did they address the correlations between the radio and optical luminosities.  Neglecting these effects when attempting to determine the distribution of radio loudness is usually given the justification that the ratio is essentially independent of cosmological model and redshift (as long as the K-corrections are the same).  The broad distribution of observed ratios obtained in this way (see Figure \ref{psir} below) deviates from a simple power law and may even have a hint of bi-modality, seemingly justifying at face value the choice of $\log R\sim1$ as the separation point between RL and RQ sources.  However, as shown in Appendix A even in the simplest cases the observed distribution (and its moments) could be very different from the intrinsic ones.  Thus, for determination of the true distribution of $R$ the data truncations must be determined and the correlations between all variables must be properly evaluated. 

Our aim in this paper is to take all these effects into account in determination of the evolution of optical and radio luminosities and their ratio and to find their distributions.  In \S \ref{dataset} we describe the data we use.  In \S \ref{simlumf} we provide an overview of the procedure used.   In \S \ref{evsec} we present our results on the correlations and evolutions of the LFs.  In \S \ref{dev} we describe the density evolution and the luminosity density evolutions, while in \S \ref{local} we calculate the LF corrected for luminosity evolution, which we call the ``local'' LF.  Finally, in \S \ref{Rdist} we evaluate the distribution of radio loudness, $R$.  This work assumes the standard $\Lambda$CDM cosmology throughout, with $H_0=71$\,km\,s$^{-1}$\,Mpc$^{-1}$, $\Omega_{\Lambda}=0.7$ and $\Omega_{m}=0.3$.

\section{Dataset} \label{dataset}

\begin{figure}
\includegraphics[width=3.5in]{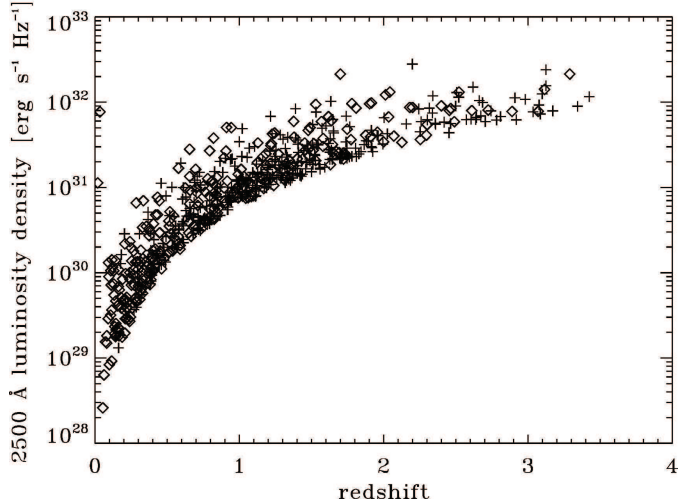}
\caption{The 2500 \AA \, rest frame absolute luminosity density for the quasars in the \citet{White00} dataset used in this analysis.  To obtain the 2500 \AA \, luminosity density we convert from observed $R$-band magnitude to flux at the integrated center band frequency, and assume an optical spectral index of 0.5 and the luminosity distance obtained from the redshift with the standard cosmology and the standard K-correction.  The crosses are the RL objects while the diamonds are the RQ.}
\label{optlums}
\end{figure}

\begin{figure}
\includegraphics[width=3.5in]{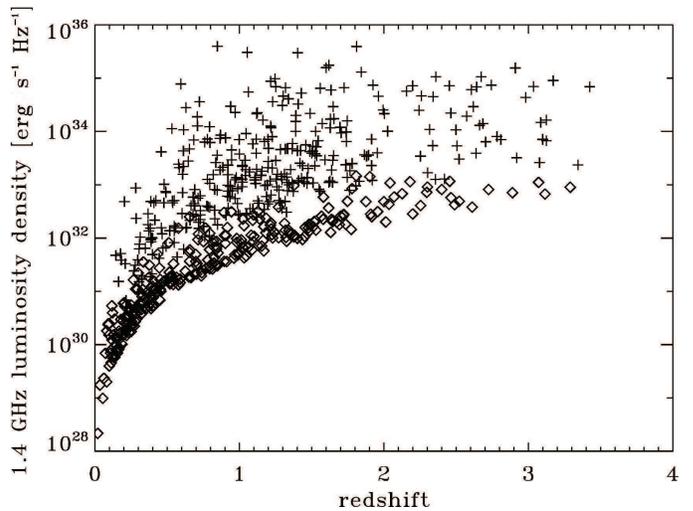}
\caption{The 1.4 GHz rest frame absolute luminosity density for the quasars in the \citet{White00} dataset used in this analysis.  To obtain the 1.4 GHz luminosity density we use the luminosity distance obtained from the redshift and the standard cosmology and the standard K-correction.  We assume an radio spectral index of 0.6.  The crosses are the RL objects while the diamonds are the RQ.}
\label{radlums}
\end{figure}

\begin{figure}
\includegraphics[width=3.5in]{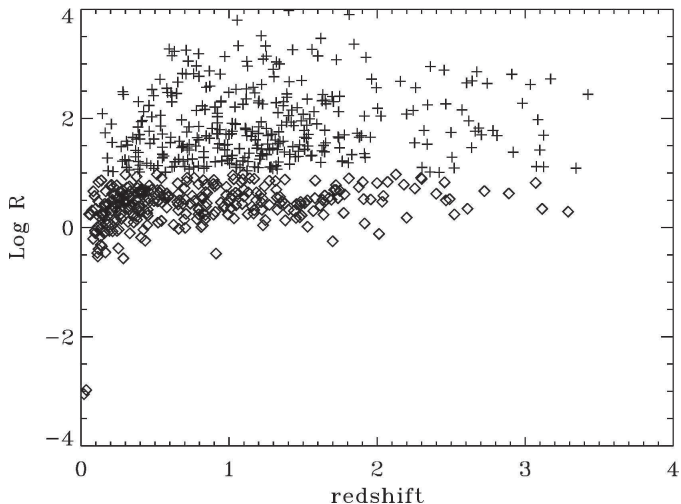}
\caption{The redshift distribution of the ratio $R$ of rest frame absolute luminosities at 5 GHz and 2500
\AA \, for the quasars in the \citet{White00} dataset used in this analysis. The 5 GHz luminosity is obtained from the 1.4 GHz luminosity assuming a radio spectral index of 0.6.  The crosses are the RL objects while the diamonds are the RQ.}
\label{sed}
\end{figure}

In order to evaluate the luminosity evolution in both radio and optical, and to separate and compare these effects, we require a data set that has both radio and optical fluxes to reasonable limits and across a range of redshifts, that contains a significant number of both RL and RQ objects.  The overlap of the FIRST bright quasar radio survey with the Automatic Plate Measuring Facility catalog of the Palomar Observatory Sky Survey (POSS-I), as presented by \citet{White00}, is such a data set.  It contains 636 objects with optical R band optical magnitudes, 1.4 GHz total and peak pixel fluxes, and spectroscopic redshifts.  The survey has a limiting R band magnitude of 17.8 or $f_{m,R}=0.22$ mJy, a limiting peak pixel 1.4 GHz flux of 1 mJy, and redshifts that range from 0.02 to 3.425.  Figures \ref{optlums} and \ref{radlums} show the radio and optical luminosities versus redshifts of the quasars in the survey, assuming the standard K-corrections for power laws with optical and radio indices $\varepsilon_{\rm opt}=0.5$ and  and $\varepsilon_{\rm rad}=0.6$ respectively (where $f_i \propto \nu^{-\varepsilon_i}$).  Figure \ref{sed} shows the radio loudness parameter $R$ versus redshift for the dataset.

This sample spans a very wide range of luminosities (5 dex in optical and 7 dex in radio) with a significant number of sources in the range  $0.1 < R < 10^4$ (with 336 RL and 300 RQ).  Therefore, it is well suited for our analysis here.  We have examined some other combined radio and optical survey data sets and found them to be not as well suited for this analysis.  For example, the combined FIRST radio survey with the 2dF optical survey as reported in \citet{Cira03} features only 12 RQ objects (of 113 total), and the combined FIRST with the Large Bright Quasar Survey (LBQS) as reported in \citet{Hewett01} has only 77 objects and different optical flux limits for the various fields, making the method employed here cumbersome.   Of course a much larger sample could be achieved combining optical data from the Sloan Digital Sky Survey (SDSS) with FIRST radio data, as in \citet{Jiang07}.  Forming such a dataset, however, would necessitate care in associating optical sources with radio ones, given the very large number of optical sources.  Issues such as how large of a radius to allow for an association and what to do about multiple matches are important considerations without immediate answers.  It is our intention to demonstrate the techniques employed here with a well established smaller dataset before moving on to a more comprehensive one involving SDSS optical data. 

\section{General remarks on correlations in luminosity functions} \label{simlumf}

The LF gives the number of objects per unit comoving volume $V$ per unit source luminosity, 
so that the number density is $dN/dV = \int dL_i \Psi_i(L_i, z)$.  To examine luminosity evolution, without loss of generality, we can write a LF in some waveband $i$ as 

\begin{equation}
\Psi_{i}\!(L_i,z) = \rho\!(z)\,\psi_i\!(L_{i}/g_{i}\!(z) , \eta_i^j)/g_i\!(z),
\label{lumeq}
\end{equation}
where $g_{i}\!(z)$ and $\rho\!(z)$ describe the luminosity evolution and comoving density evolution with redshift respectively and $\eta_i^j$ stands for parameters that describe the shape (e.g. power law indices and break values) of the $i$ band LF (we use the normalization $\int_0^{\infty}  \psi_i(L_i)dL_i = 1$).\footnote{There are in principle other possible parameters, e.g. the spectral indices.  We can ignore them for the purposes of the analysis here on the assumption that they either do not evolve strongly with redshift {\it or} are not strongly correlated with any of the luminosities in question.}  In what follows we assume a non-evolving shape for the LF (i.e. $\eta_i^j=$ const, independent of $L$ and $z$), which is a good approximation for determining the global evolutions.  Once these are determined this hypothesis can be tested and results amended.  However, for more complicated functional forms with variable $\eta_i^j$, e.g. for luminosity dependent density evolution, the determination of the variations will require a large sample with significant numbers of objects in reasonably narrow redshift and luminosity bins.

Given this assumption then once the luminosity evolution $g_{i}\!(z)$ is calculated, the density evolution $\rho\!(z)$ and local LF $\psi_{i}\!(L_{i}') \equiv \psi_{i}\!(L_i/g_{i}(z))/g_i\!(z)$ can be determined.\footnote{The method developed by EP that we shall use below actually gives the cumulative functions $\sigma(<z)=\int_0^z{\rho}(z')\,[dV(z')/dz']\,dz'$ and $\phi(>L')=\int_{L'}^\infty \psi(L'')\,dL''$. The differential functions $\rho$ and $\psi$ are obtained by differentiation.}  We consider this form of the LF for luminosities in different bands, allowing for separate (optical and radio) luminosity evolution.    

1. As is often done, one might naively assume that the joint LF $\Psi\!(L_{\rm opt},L_{\rm rad},z)$ is separable into two forms like equation \ref{lumeq} with a common density evolution.  However, as discussed in \S \ref{intro}, because the optical and radio luminosities of the quasars are, in general, highly correlated, the simultaneous determination of the LFs of both requires care.  The first step in this procedure should be to determine the degree and form of the correlation between the optical and radio luminosities.  As described below, the EP method allows us to determine whether any pair of variables are independent or correlated.  Once it is determined that they are correlated one should seek a coordinate transformation to define a new pair of variables which are independent.  This requires a parametric form for the transformation.  One can define a new luminosity which is a combination of the two; we can define a ``correlation reduced radio luminosity'' $L_{\rm crr}=L_{\rm rad} / F(L_{\rm opt}/L_{fid}),$ where the function $F$ describes the correlation between $L_{\rm rad}$ and $L_{\rm opt}$ and $L_{fid}$ is a fiducial luminosity taken here to be $10^{28}$\,erg\,sec$^{-1}$\,Hz$^{-1}$.  This is a convenient choice for $L_{fid}$ as it is lower than the lowest 2500 $\AA \,$ luminosity considered in our sample, but results do not depend on the particular choice of numerical value.  For the correlation function we will assume a simple power law 
\begin{equation}
L_{\rm crr} = {{L_{\rm rad}} \over {(L_{\rm opt}/L_{fid})^{\alpha}}}
\label{rcrdef}
\end{equation}
where $\alpha$ is a bulk power law correlation index to be determined by a fit to the data.  This is essentially a coordinate rotation in the log-log luminosity space.  As shown in \S \ref{evsec} below, EP also prescribe a method to determine a best fit value for the index $\alpha$ which orthogonalizes the new luminosities.  Given the correlation function we can then transform the data (and its truncation) into the new independent pair of luminosities $(L_{\rm opt}$ and $L_{\rm crr})$, whose distribution can be represented as

\begin{eqnarray}
\Psi\!(L_{\rm opt},L_{\rm crr},z) = 
\nonumber \\ \rho(z) \, \times \, \psi_{\rm opt}(L_{\rm opt}/g_{\rm opt}, \eta_{\rm opt}^j)/g_{\rm opt} \nonumber \\ \times \, \psi_{\rm crr}(L_{\rm crr}/g_{\rm crr}, \eta_{\rm crr}^j)/g_{\rm crr}.
\label{fun}
\end{eqnarray} 

2. The next step is determination of the two {\it independent}  luminosity-redshift correlation functions $g_{\rm opt}$ and $g_{\rm crr}$ which describe the luminosity evolutions.  The procedure for determination of these functions is similar to the ones for removing the correlations between the luminosities except now we make coordinate transformations in the $L_{\rm opt}-z$ and $L_{\rm crr}-z$ spaces.  We assume simple forms
\begin{equation}
g_i(z)=(1+z)^{k_i}
\label{Levolution}
\end{equation}
so that $L'_i=L_i/g_i(z)$ refer to the local ($z=0$) luminosities.\footnote{This is an arbitrary choice. One can chose any other fiducial redshift by defining $g_i(z)=[(1+z)/(1+z_{fid})]^{k_i}$.}  The full procedure is detailed in \S \ref{evsec}.

3. The density evolution function $\rho(z)$ is determined by the method shown in EP (see \S \ref{dev} below).  Once all correlations are removed we end up with a local separable LF as in equation \ref{fun}.

4. The local LFs of uncorrelated luminosities $L'_{\rm opt}$ and $L'_{\rm crr}$ can then be used to recover the local radio LF by a straight forward integration over $L'_{\rm crr}$ and the true local optical LF as 

\begin{eqnarray}
\psi_{\rm rad}\!(L_{\rm rad}') = 
\nonumber \\ \int_0^{\infty} { \psi_{\rm opt}\!(L_{\rm opt}') \, \psi_{\rm crr}\left({{ L_{\rm rad}' } \over ({{L_{\rm opt}'/L_{fid}})^{\alpha} } } \right) \, {{dL_{\rm opt}'} \over ({{L_{\rm opt}'/L_{fid}})^{\alpha}}  } \,}
\label{localrad}
\end{eqnarray}
As stated above this procedure can be used for the determination of the radio LF at any redshift, from which one can deduce that the radio luminosities also undergo luminosity evolution with 
\begin{equation}
g_{\rm rad}\!(z) = g_{\rm crr}\!(z) \, \times \, [g_{\rm opt}\!(z)]^{\alpha}
\label{gradform}
\end{equation}
(cf equation \ref{rcrdef})

5. Similarly we can determine the local distribution of the radio to optical luminosity ratio, 
$R' = L'_{\rm rad}/L'_{\rm opt} = L'_{\rm crr} \, \times \, {L'_{\rm opt}}^{\alpha-1} \, \times {L_{fid}}^{-\alpha}$, as

\begin{equation}
G_{R'} = \int_0^{\infty} { \psi_{\rm opt}\!(L_{\rm opt}') \, \psi_{\rm crr}\left({{ R' \, L_{fid}} \over ({{L_{\rm opt}'/L_{fid}})^{\alpha-1} } } \right) \, {{dL_{\rm opt}'} \over {{L_{\rm opt}'}^{\alpha - 1} \, L_{fid}}  } \,}
\label{localr}
\end{equation}
and its evolution 
\begin{equation}
g_{R}\!(z) = g_{\rm crr}\!(z) \, \times \, [g_{\rm opt}\!(z)]^{\alpha-1} = {g_{\rm rad} \over g_{\rm opt}}
\label{Rexp}
\end{equation}

\section{Correlation functions} \label{evsec}

We now describe results obtained from the use of the procedures described in \S \ref{simlumf} on the data described in \S \ref{dataset}.  Here we first give a brief summary of the algebra involved in the EP method.  We follow closely the steps described in \citet{MP99}.  This method uses the Spearman rank test to determine the best-fit values of parameters describing the correlation functions using  the test statistic 

\begin{equation}
\tau = {{\sum_{j}{(\mathcal{R}_j-\mathcal{E}_j)}} \over {\sqrt{\sum_j{\mathcal{V}_j}}}}
\label{tauen}
\end{equation}
to test the independence of two variables in a data set, say ($x_j,y_j$) for  $j=1, \dots, n$.  Here $R_j$ is the $y$ rank of the data point $j$ in a set associated with it.  For a untruncated data (i.e. data truncated parallel to the axes) the {\it associated set} of point $j$ includes all of the  $x_k < x_j$.  If the data is truncated one must form the {\it associated set} consisting only of those points of lower $x$ value that would have been observed if they were at the $x$ value of point $j$ given the truncation.  As an example, if we have one sided truncations as in Figures \ref{optlums} and \ref{radlums}, then the associated set $A_j \, = \, \{ \, k:y_k \,> \, y_j, \, y^-_k\,<\,y_j \, \}$, where $y^-_k$ is the limiting $y$ value of data point $i$ (see EP for a full discussion of this method).  

If ($x_j,y_j$) were independent then the rank $\mathcal{R}_j$ should be distributed uniformly between 0 and 1 with the expectation value and variance $\mathcal{E}_j=(1/2)(j+1)$ and $\mathcal{V}_j=(1/12)(j^{2}+1)$, respectively.  Independence is rejected at the $n \, \sigma$ level if $\vert \, \tau \, \vert > n$.  To find the best fit correlation the $y$ data are then adjusted by defining $y'_j=y_j/F(x_j)$  and the rank test is repeated, with different values of parameters of the function $F$.

\subsection{Radio-Optical Luminosity Correlation} \label{roevsec}

\begin{figure}
\includegraphics[width=3.5in]{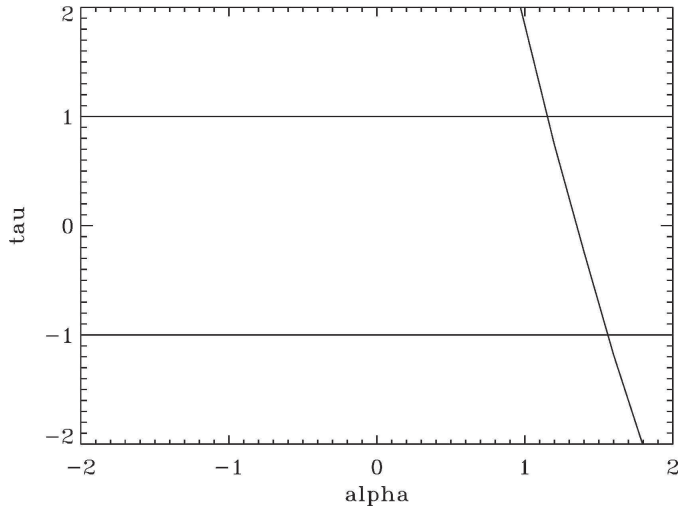}
\caption{The value of the $\tau$ statistic as given by equation \ref{tauen} as a function of $\alpha$ for the relation $L_{\rm rad} \propto (L_{\rm opt})^{\alpha}$, where $L_{\rm opt}$ and $L_{\rm rad}$ are the optical and radio luminosities, respectively, for the quasars in the dataset.  The 1 $\sigma$ range for the best fit value of $\alpha$ is where $\vert \, \tau \, \vert \leq 1$.  It is seen that the observed optical and radio luminosities are strongly positively correlated, with a linear or slightly higher power law relation. } 
\label{taus}
\end{figure}

The radio and optical luminosities are obtained from radio and optical fluxes from a two flux limited sample so that the data points in the two dimensional flux space are truncated parallel to the axes which we consider to be untruncated.  Since the two luminosities have essentially the same relationship with their respective fluxes, except for a minor difference in the K-correction terms, we can consider the luminosity data to also be untruncated.  In that case as mentioned above the determination of the associated set is trivial and one is dealing with the standard Spearman rank test.  Assuming the correlation function between the luminosities $F(x)=x^{\alpha}$ we calculate the test statistic $\tau$ as a function of  $\alpha$.  Figure \ref{taus} shows the absolute value of the $\tau$ vs $\alpha$, from which we get the best fit value of $\alpha=1.3$ with one $\sigma$ range $\pm 0.2$.  As expected $\alpha$ is near unity, and the value $\alpha$=1 is not ruled out with a high significance.  As discussed in Appendix B, this correlation may be inherent in the population or may be an artifact of the flux limits and wide range of redshifts, although this particular point is not important for the analysis going forward, as the rotation to $L_{\rm crr}$ is a technique to achieve independent variables ($L_{\rm opt}$ and $L_{\rm crr}$) in the context of the data present to recover the inherent redshift evolutions.

\subsection{Luminosity-Redshift Correlations} \label{method}

We now describe our results on determination of the luminosity evolution, i.e.  the luminosity-redshift correlation functions $g_i\!(z)$, which according to equation \ref{Levolution} reduces to determination of the values of the indices $k_i$.  The basic method for determining the best fit $k_i$ is the same as above but in this case the procedure is more complicated for several reasons.  First, as evident from Figures \ref{optlums} and \ref{radlums} the $L_i-z$ data are heavily truncated due to the flux limits.  Second, we now are dealing with a three dimensional distribution  ($L_{\rm crr}, L_{\rm opt}, z$) and two correlation functions [$g_{\rm crr}\!(z)$ and $g_{\rm opt}\!(z)$].

Specifically, since we have two criteria for truncation, the associated set for each object includes only those objects that are sufficiently luminous in both bands to exceed {\it both} flux minima for inclusion in the survey if they were located at the redshift of the object in question.  Consequently, we have a two dimensional minimization problem, because both the optical and correlation reduced radio evolution factors, $g_{\rm opt}\!(z)=\!(1+z)^{k_{\rm opt}}$ and $g_{\rm crr}\!(z)=\!(1+z)^{k_{\rm crr}}$, come into play, as the luminosity cutoff limits for a given redshift are adjusted by powers of $k_{\rm opt}$ and $k_{\rm rad}$ too.  

\begin{figure}
\includegraphics[width=3.5in]{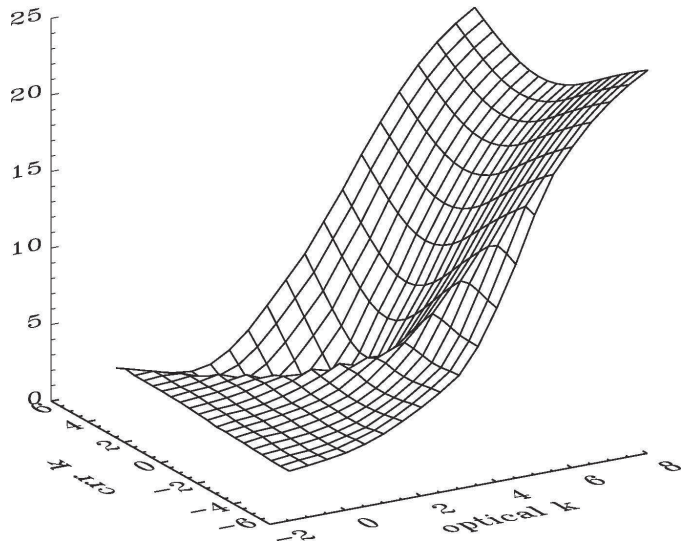}
\caption{Surface plot of the value of $\tau_{comb}$ for the dataset as a whole showing the location of the minimum region where the favored values of $k_{\rm opt}$ and $k_{\rm crr}$ lie.}
\label{tausss}
\end{figure}

We form a test statistic $\tau_{comb} = \sqrt{\tau_{\rm opt}^2 + \tau_{\rm crr}^2} $ where $\tau_{\rm opt}$ and $\tau_{\rm crr}$ are those evaluated considering the objects' optical and correlation reduced radio luminosities, respectively.  The favored values of $k_{\rm opt}$ and $k_{\rm crr}$ are those that simultaneously give the lowest $\tau_{comb}$ and, again, we take the $1 \sigma$ limits as those in which $\tau_{comb} \, < 1$.  For visualization, Figure \ref{tausss} shows a surface plot of $\tau_{comb}$

We have verified this method with a simulated Monte Carlo data set in which objects are distributed in redshift and given randomized luminosities in accordance with set optical and radio evolutions.  The algorithm can recover the evolutions correctly provided that they aren't wildly different, i.e. one very positive and the other very negative.  

\begin{figure}
\includegraphics[width=3.5in]{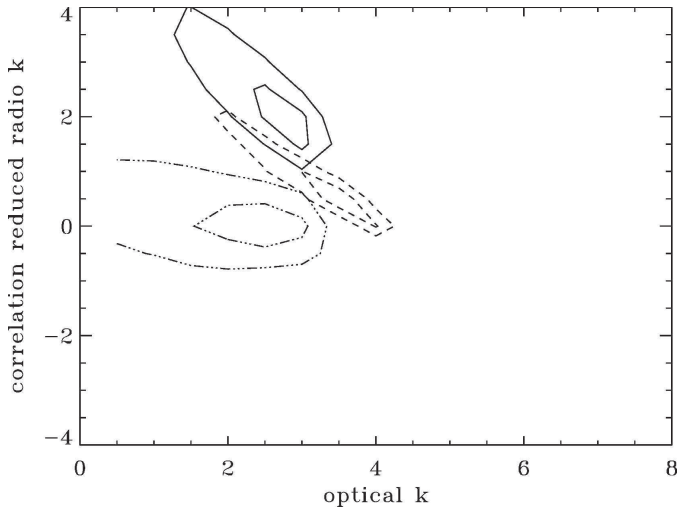}
\caption{The $1 \sigma$ and $2 \sigma$ contours for the simultaneous values of $k_{\rm opt}$ and $k_{\rm crr}$ where the optical and correlation reduced radio luminosity evolutions are $g_{\rm opt}\!(z)=\!(1+z)^{k_{\rm opt}}$ and $g_{\rm crr}\!(z)= (1+z)^{k_{\rm crr}}$.  The best fit radio luminosity evolution can be reconstructed from $g_{\rm rad} = g_{\rm crr} \, \times \, {g_{\rm opt}}^{1.3}$.  Results are shown for the data set evaluated as a whole (solid contours), and for the RL (dash-dot contours) and RQ (dashed contours) populations evaluated separately.  It is evident that splitting the population before determination of the luminosity evolutions introduces a bias into the determinations, as discussed in \S \ref{method}. }
\label{alphas}
\end{figure}

Figure \ref{alphas} shows the best fit values of $k_{\rm opt}$ and $k_{\rm crr}$ and taking the1 and 2 $\sigma$ contours.   Results are shown for the entire dataset taken as a whole, and also with the data split into the RL and RQ subsets.  The radio luminosity evolution itself can be recovered by equation \ref{gradform}.

Given the tight constraints achieved when the dataset is considered as a whole, and the sharp bifurcation when the set is split into the RL and RQ populations, it is evident that splitting the population before determination of the luminosity evolutions introduces a bias into the determinations.  This is expected because differing evolutions will have a strong effect on the likelihood that objects at a given redshift will fall on the RL or RQ side of the division according to the standard definition used here, and the data for each set will be artificially truncated along $R=10$ as a function of the evolutions.

We see that positive evolution in both radio and optical wavebands is favored.  The minimum value of $\tau_{comb}$ favors an optical evolution of $k_{\rm opt}$ = 3.0 and a radio evolution of $k_{\rm rad}$ = 5.4, but uncertainty at the $1 \sigma$ level allows the range of $k_{\rm opt}$ from 2.5 to 3.25, and $k_{\rm rad}$ from 5.3 to 5.75.  Therefore, we conclude that quasars have undergone a significantly greater radio evolution relative to optical evolution with redshift.  In the above analysis we have assumed  sharp truncation boundaries and that the data is complete above the boundaries.  As discussed in \S \ref{testass} this may not be the case for the FIRST radio data.  If an estimate of the uncertainty from the consideration of possible radio incompleteness at faint fluxes is included, the favored optical range enlarges to $k_{\rm opt}$ from 1.25 to 3.75, with a slightly lower best fit value of $k_{\rm opt}$ = 2.0. Due to combined effects on $k_{\rm opt}$ and $k_{\rm crr}$, the value of $k_{\rm rad}$ is not much affected by allowing for possible radio incompleteness at faint fluxes, perhaps counterintuitively.

\section{Density evolution} \label{dev}

Next we determine the density evolution $\rho\!(z)$.  One can define the cumulative density function 

\begin{equation}
\sigma\!(z) = \int_0^z {\rho\!(z) \, dz}
\end{equation}
which, following \citet{P92} based on \citet{L-B71}, can be calculated by

\begin{equation}
\sigma\!(z) = \prod_{j}{(1 + {1 \over m\!(j)})}
\end{equation}
where $j$ runs over all objects with a redshift lower than or equal to $z$, and $m(j)$ is the number of objects with a redshift lower than the redshift of object $j$ {\it which are in object j's associated set}.  In this case, the associated set is again those objects with sufficient optical and radio luminosity that they would be seen if they were at object $j$'s redshift.  The use of only the associated set for each object removes the biases introduced by the data truncation.  Then the density evolution $\rho\!(z)$ is 

\begin{equation}
\rho\!(z) = {d \sigma\!(z) \over dz}
\label{rhoeqn}
\end{equation}

However, to determine the density evolution, the previously determined (in \S \ref{evsec}) luminosity evolution must be taken out.  Thus, the objects' optical and radio luminosities, as well as the optical and radio luminosity limits for inclusion in the associated set for given redshifts, are scaled by taking out factors of $g_{\rm opt}\!(z)=\!(1+z)^{k_{\rm opt}}$ and $g_{\rm rad}\!(z)=\!(1+z)^{k_{\rm rad}}$, with $k_{\rm opt}$ and $k_{\rm rad}$ determined as above.    

\begin{figure}
\includegraphics[width=3.5in]{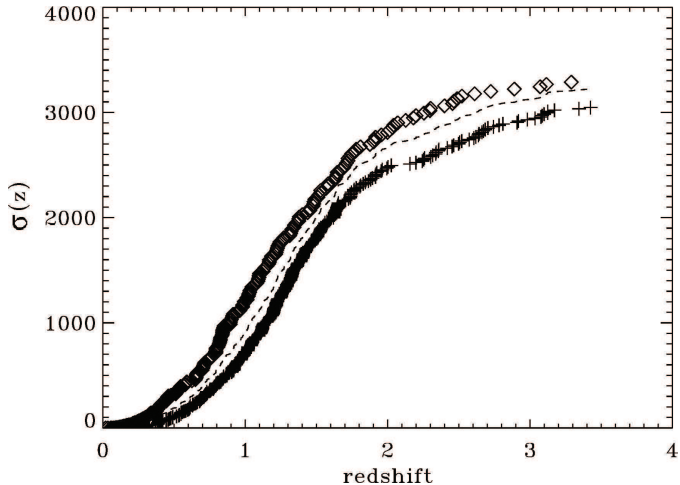}
\caption{The cumulative density function $\sigma\!(z)$ vs. redshift for the RL (crosses) RQ (diamonds) and all (dashed line) quasars in the data set.  The normalization of $\sigma\!(z)$ is arbitrary, and the RL data has been shifted vertically for clarity.  A piecewise quadratic fit to $\sigma\!(z)$ is used to determine $\rho\!(z)$ by equation \ref{rhoeqn}. }
\label{sigma}
\end{figure}

\begin{figure}
\includegraphics[width=3.5in]{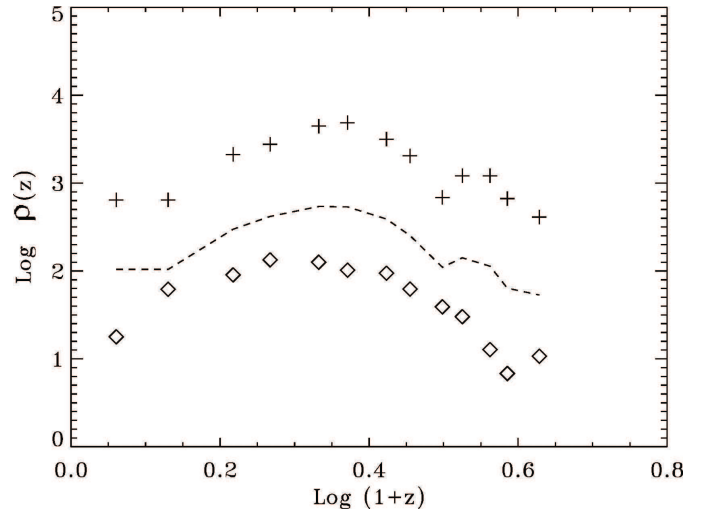}
\caption{The density evolution $\rho\!(z)$ vs. redshift for the for the RL (crosses) RQ (diamonds) and all (dashed line) quasars in the data set, shown with customary log scales.  The normalization of $\rho\!(z)$ is arbitrary and the curves have been shifted vertically for clarity. }
\label{rholog}
\end{figure}

Figures \ref{sigma} and \ref{rholog} show $\sigma\!(z)$ and $\rho\!(z)$ for the objects in the data set.  We evaluate and display the density evolution separately for the RL and RQ objects and for the dataset as a whole to compare them.  It is seen that the two groups, divided in this way, exhibit very similar density evolution.  The number density of quasars seems to peak at between redshifts 1 and 1.5, a little earlier than generally thought for the most luminous quasars \citep[e.g.][]{Shaver96}, and earlier than that found in \citet{Richards06}, but similar to the peak found for less luminous quasars by \citet{Hopkins07}, and in agreement with \citet{MP99}.  

\begin{figure}
\includegraphics[width=3.5in]{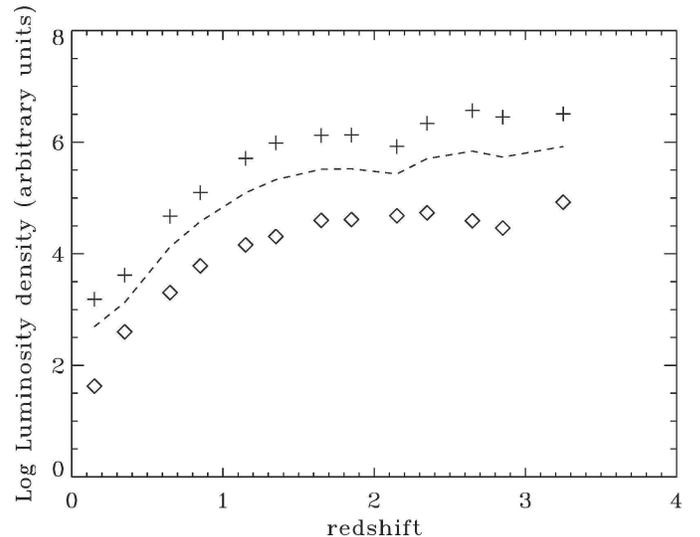}
\caption{The radio luminosity density function $\pounds_{\rm rad}\!(z)$ vs. redshift for the for the RL (crosses) RQ (diamonds) and all (dashed line) quasars in the data set.  The normalization of $\pounds_{\rm rad}\!(z)$ is arbitrary, and the values have been shifted vertically for clarity.  It is seen that the two populations have very similar luminosity density evolution with redshift. }
\label{LDFrad}
\end{figure}

Knowing both the luminosity evolutions $g_{i}\!(z)$, and the density evolution $\rho\!(z)$, one can form the luminosity density functions $\pounds_{i}\!(z)$, which are the total rate of production of energy of quasars as a function of redshift.  We show this for the radio luminosity density $\pounds_{\rm rad}\!(z)$.  As evident the two populations of RL and RQ have very similarly shaped radio luminosity density functions (Figure \ref{LDFrad})

\section{Local luminosity functions} \label{local}

\subsection{General Considerations}

In a parallel procedure we can use the `local' (redshift evolution taken out, or 'de-evolved') luminosity $L'_i$ distributions (and de-evolved luminosity thresholds) to determine  the `local' LFs $\psi_{i}\!({L_i}')$, where again the $i$ represents the waveband, and the prime indicates that the luminosity evolution has been taken out.  We first obtain the cumulative LF 

\begin{equation}
\Phi_i\!(L_{i}') = \int_{L_{i}'}^{\infty} {\psi_i\!(L_{i}'') \, dL_{i}''}
\end{equation}
and, following \citet{P92}, $\Phi_i\!(L_{i}')$ can be calculated by

\begin{equation}
\Phi_i\!(L_{i}') = \prod_{k}{(1 + {1 \over n\!(k)})}
\label{phieq}
\end{equation}
where $k$ runs over all objects with a luminosity greater than or equal to $L_i$, and $n(k)$ is the number of objects with a luminosity higher than the luminosity of object $k$ which are in object $k$'s associated set, determined in the same manner as above. 
The LF $\psi_i\!(L_{i}')$ is 

\begin{equation}
\psi_i\!(L_{i}') = - {d \Phi_i\!(L_{i}') \over dL_{i}'}
\label{psieqn}
\end{equation}

In \S \ref{evsec} we have determined the luminosity evolution for two independent functions, the optical luminosity $L_{\rm opt}$ and the correlation reduced radio luminosity $L_{\rm crr}$.   We can form the local optical $\psi_{\rm opt}\!(L_{\rm opt}')$ and correlation reduced radio $\psi_{\rm crr}\!(L_{\rm crr}')$ LFs straightforwardly, by taking the evolutions out.  As before, the objects' luminosities, as well as the luminosity limits for inclusion in the associated set for given redshifts, are scaled by taking out factors of $g_{\rm crr}\!(z)=\!(1+z)^{k_{\rm crr}}$ and $g_{\rm opt}\!(z)=\!(1+z)^{k_{\rm opt}}$, with $k_{\rm crr}$ and $k_{\rm opt}$determined in \S \ref{evsec}.  We use the notation $L \rightarrow L' \equiv L/g(z) $.

\subsection{Local optical luminosity function}\label{locopt}

\begin{figure}
\includegraphics[width=3.5in]{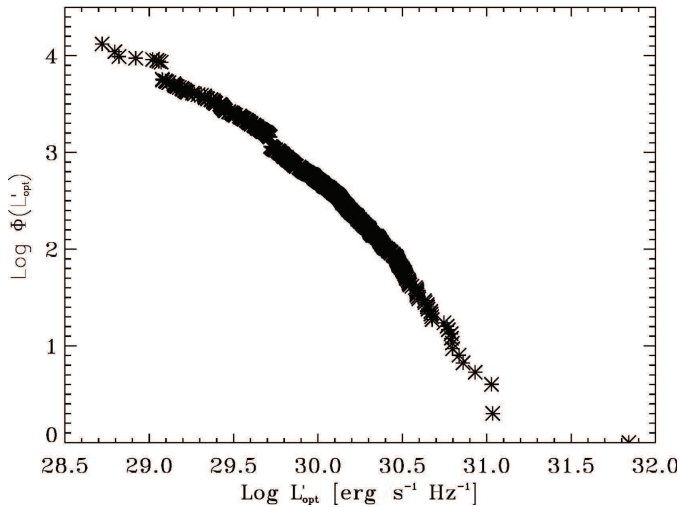}
\caption{The cumulative local optical LF $\Phi_{\rm opt}\!(L_{\rm opt}')$ for the quasars in the data set.  A piecewise quadratic fit to $\Phi\!(L_{\rm opt}')$ is used to determine $\psi_{\rm opt}\!(L_{\rm opt}')$ by equation \ref{psieqn}. The normalization of $\Phi_{\rm opt}\!(L_{\rm opt}')$ here is arbitrary.}
\label{phiopt}
\end{figure}

\begin{figure}
\includegraphics[width=3.5in]{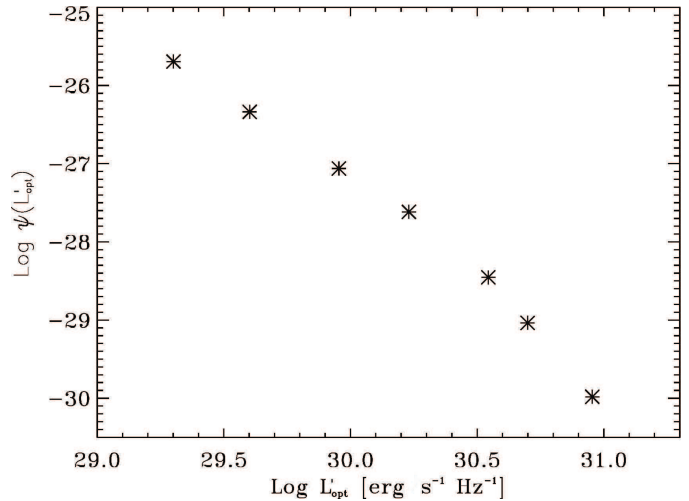}
\caption{The local optical LF $\psi_{\rm opt}\!(L_{\rm opt}')$ for the quasars in the data set.  The normalization of $\psi_{\rm opt}\!(L_{\rm opt}')$ here is arbitrary.  }
\label{psiopt}
\end{figure}

\begin{figure}
\includegraphics[width=3.5in]{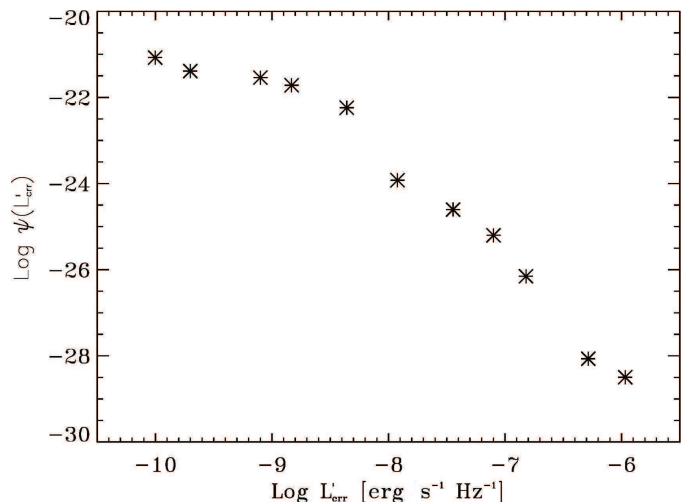}
\caption{The local correlation reduced radio LF $\psi_{\rm crr}\!(L_{\rm crr}')$ for the quasars in the data set. The normalization of $\psi_{\rm crr}\!(L_{\rm crr}')$ here is arbitrary. For clarity, as plotted here, we have taken a numerical factor of $(L_{\rm fid})^{\alpha}$ out of $L_{\rm crr}’$ (cf Equation \ref{rcrdef}). }
\label{psircr}
\end{figure}

Figures \ref{phiopt} and \ref{psiopt} show the local cumulative $\Phi_{\rm opt}\!(L_{\rm opt}')$ and differential $\psi_{\rm opt}\!(L_{\rm opt}')$ local optical LFs of the quasars in the \cite{White00} dataset, while figure \ref{psircr} shows the local correlation reduced radio LF, $\psi_{\rm crr}\!(L_{\rm crr}')$.  

The optical LF shows evidence of a break at $2 \times 10^{30}$\,erg\,sec$^{-1}$\,Hz$^{-1}$, which was present already in data used in \citet{P73}.  Fitting a broken power law yields values $-2.0\pm0.2$ and $-3.2\pm0.2$ below and above the break, respectively.  If we allow for the possibility of additional uncertainty resulting from the consideration of possible radio incompleteness at faint fluxes (see discussion in \S \ref{testass}), the range on the power law above the break increases to $-2.8\pm0.4$.   As the optical LF has been studied extensively in various AGN surveys, we can compare the slope of $\psi_{\rm opt}\!(L_{\rm opt}')$ obtained here to values reported in the literature.  For example, \citet{Boyle00}, using the 2dF optical data set (but with no radio overlap criteria) use a customary broken power law form for the LF, with values ranging from $-$1.39 to $-$3.95 for different realizations, showing reasonable agreement.\footnote{It should be noted that they parameterize evolution differently and work in absolute magnitudes rather than luminosities, however the slopes of their fits to the LF as they parameterize it are applicable, as can be seen in their section 3.2.2.}

\subsection{Local radio luminosity function}

\begin{figure}
\includegraphics[width=3.5in]{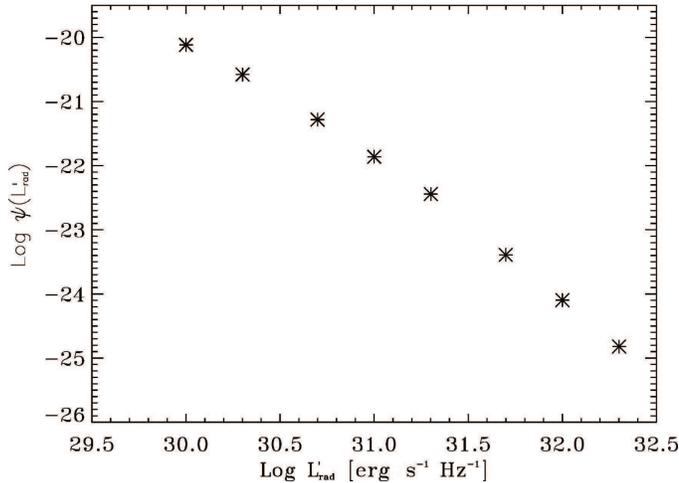}
\caption{The local radio LF $\psi_{\rm rad}\!(L_{\rm rad}')$ for the quasars in the data set.  The normalization of $\psi_{\rm rad}\!(L_{\rm rad}')$ here is arbitrary.}
\label{psirad}
\end{figure}

With $\psi_{\rm opt}\!(L_{\rm opt}')$ and $\psi_{\rm crr}\!(L_{\rm crr}')$, we can determine the local radio LF $\psi_{\rm rad}\!(L_{\rm rad}')$  with equation \ref{localrad}.  Figure \ref{psirad} shows the local radio LF $\psi_{\rm rad}\!(L_{\rm rad}')$ reconstructed in this way.  It is seen that the local radio LF contains a possible break around $10^{31}$\,erg\,sec$^{-1}$\,Hz$^{-1}$, with a power law slope of $-1.7\pm0.1$ below the break and $-2.4\pm0.1$ above it.   These ranges for the power law above the break are increased slightly to $-2.2\pm0.3$ if the effects of possible radio incompleteness are included, as in \S \ref{testass}.  The slope above the break seen here is similar to earlier results of \citet{Schmidt72} and \citet{P73} which probed only those luminosities.  A more complete comparison can be made with \citet{MS07}, who form radio LFs of local sources in the Second Incremental Data Release of the 6 degree Field Galaxy Survey (6dFGS) radio catalog.  For the sources they identify as AGN, they find a break at $3.1 \times 10^{31}$\,erg\,sec$^{-1}$\,Hz$^{-1}$, with slopes of $-2.27\pm0.18$ and $-1.49\pm0.04$ above and below the break (converting to luminosity units).

\section{Distribution of radio loudness ratios}  \label{Rdist}

\begin{figure}
\includegraphics[width=3.5in]{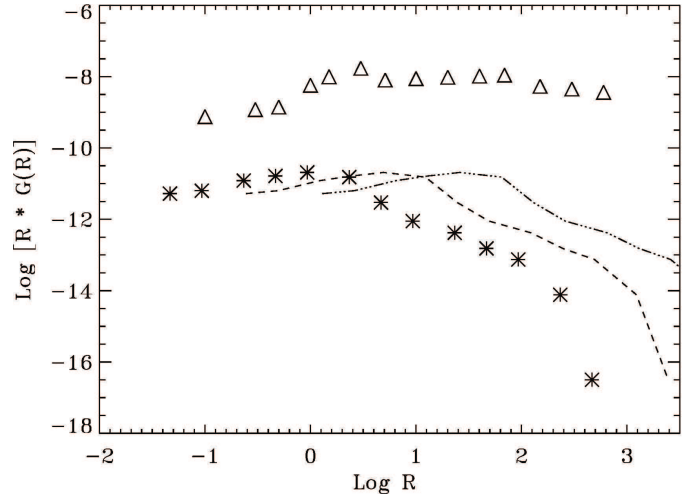}
\caption{The local distribution $G_{R}\!(R)$ in the 5 GHz radio to 2500 \AA \, optical luminosity ratio R, plotted as $R\, \times \, G_{R}\!(R)$, for the quasars in the data set.  The stars are from $G_{R'}\!(R')$  as determined by the method of Equation \ref{localr}, taking account of the truncations and correlations in the luminosity evolutions, while the triangles result from forming a distribution with a naive use of the objects' raw ratio.  The normalization is arbitrary and the curves have been shifted vertically for clarity.  It is seen that the naive method gives a hint of a bi-modal distribution, while the proper method does not.  Also shown is the proper radio loudness distribution $G_R\!(R,z)$ at redshifts z=1 (dashed line) and z=3 (dash-dot line), evolved according to the form of equation \ref{Rexp}. }
\label{psir}
\end{figure}

As stated in the introduction, naively one may expect that because the ratio $R$ is independent of cosmological model and nearly independent of redshift, the raw observed distribution would provide a good representation of the true distribution of this ratio. In Figure \ref{psir} we show this raw distribution by the triangles, arrived at by using the raw values of $R$ from the data and forming a distribution in the manner of equations \ref{phieq} and \ref{psieqn} with no data truncations.  It appears that this naive approach shows a hint of possible bi-modality with ${\rm \log} R\sim1$ as the dividing value.\footnote{We note that in general apparent bi-modalities often do not stand up to rigorous statistical tests.}  

As discussed in \S \ref{simlumf}, we can reconstruct the local distribution of $G_{R'}\!(R')$, as in Equation \ref{localr}, which provides for a more proper accounting of the biases and truncations.  The results of this calculation are also shown in Figure \ref{psir}.  The distribution calculated in this way clearly is different than the raw distribution, and does not show any apparent bi-modality.  There is still a possible feature in the same region (${\rm \log} R\sim1$) where the raw distribution shows a dip.  This feature is of marginal significance and results from the similarly shaped feature in $\psi_{\rm crr}\!(L_{\rm crr}')$ centered around $L_{\rm crr}' = 10^{-8}$\,erg\,sec$^{-1}$\,Hz$^{-1}$.  Even if significant, this change in slope cannot be taken as evidence for two physically distinct populations, but could be a useful point to make an arbitrary division into RL and RQ objects.  

We also know that the best fit redshift evolution of the ratio, given equation \ref{Rexp}, is $g_R\!(z) = (1+z)^{2.6}$.  The change in the distribution of $R$ with increasing redshift is also shown in Figure \ref{psir}.\footnote{Note that we have not included the density evolution which will shift the curves vertically but not change their shape.}  Another way to look at this is that we have found that the radio luminosity evolves at a different rate than the optical luminosity, with the consequence that their ratio is a function of redshift.  The radio loudness of the population increases by a factor of 5 by redshift 1, and by a factor of 28 by redshift 3.  This is in disagreement with the result presented by \citet{Jiang07} who show a decrease in fraction of RL sources with increasing redshift, which could be the case if the radio luminosity were to evolve more slowly than the optical luminosity. They however do not determine individual evolutions or LFs.  On the other hand, \citet{Miller90}  have noted that the fraction of RL quasars may increase with redshift, which they attribute to a difference in the evolutions of the two populations (RL and RQ).  \citet{Donoso09} compute radio and optical LFs at different redshifts and reach the same conclusion.  \citet{Cira06} also find that the radio loud fraction may modestly increase at high redshift.  Although not directly comparable, \citet{LaFranca10} show a similar evolution for $R_x$, the ratio of radio to X-ray luminosity, as we show here for $R$.  We note that our results favor one population, in the sense that the distribution of $G(R)$, recovered from considering the data truncations inherent in the survey and correlations between the luminosities, is continuous. 

\section{Tests of assumptions}\label{testass}

{\it Power law parameterization}: One may raise the concern that the simple power law parameterization used for the redshift luminosity evolutions (equation \ref{Levolution}) may not be the most ideal one.  In particular, it may not accurately represent the evolutions at the highest redshifts considered here.  To check this, we repeat the analysis with a different parameterization for the luminosity evolution which allows for a flattening at higher redshifts,

\begin{equation}
g_i(z)={ {(1+z)^{k_i}} \over {1+({{1+z} \over {4}})^{k_i}} },
\label{altevolution}
\end{equation}
where $i$ again represents the optical or correlation reduced radio luminosity.  In this parameterization, the functional form for the radio luminosity evolution $g_{\rm rad}(z)$ and the evolution of the radio loudness parameter $g_R(z)$ are lengthier expressions involving $k_{\rm opt}$ $k_{\rm crr}$ and $\alpha$, given equations \ref{gradform} and \ref{Rexp}.

This alternate parameterization for the evolutions does not appreciably effect the results.  The best fit evolution factors as a function of redshift under the alternate parameterization differ very little from those in the simple parameterization to redshift 3.5 (the highest object in the sample).  In a data set with higher redshift objects, the form of the parameterization will be more consequential.

{\it Luminosity dependent density evolution}:  Another concern may be that luminosity dependent density evolution (LDDE), which is not considered in the functional forms for the LF used here, may more accurately represent the evolution of the LF.  As a check of this effect, we divide the data into high and low luminosity halves (cutting on optical luminosity), and check the similarity of the computed density evolutions for the two sets versus that computed assuming the absence of LDDE.  Given that an artificial difference is already introduced in the two halves because there are a lack of low luminosity objects in the high redshift sample and a lack of high luminosity redshifts in the low redshift sample (see Figures \ref{optlums} and \ref{radlums}), we conclude that the density evolutions determined in this way are sufficiently similar to justify neglecting LDDE.

{\it Optical measurement errors}:  There is the possibility that errors in the optical magnitudes could lead to a bias which could affect the results.  The bias introduced by these errors would be negligible if the LogN-LogS was flat (i.e. $N(>S) \propto S^0$ and $dN/dS \propto S^{-1}$). Since the number density of sources increases with decreasing flux, it is more likely that a source will be included than excluded.  However, the magnitude of this effect will depend on the faint end source counts slope, and the shallower the slope the smaller the effect.  The magnitude of this effect is proportional to $[(1+m_{below}) \, \delta]^2$ where $\delta$ is the fractional error in flux and $m_{below}$ is the faint end differential source counts slope (i.e. $dN/dS \propto S^{m_{below}}$).  For the reported POSS-I magnitude errors of 0.2, for the faint end magnitudes of $\sim$17.8, $\delta$ is less than 0.2.  For $m_{below} \sim-2$ and $\delta \sim 0.2$ the bias will be less than 4\% averaged over the faintest fluxes, which would be manifest in raising the faint end cumulative source counts slope by an amount still considerably smaller than this, as it is fit over a larger range of fluxes.  Previously, \citet{CP93} investigated this effect for a flux limited dataset and showed that the difference was minor between using a Gaussian distribution of fluxes for each source instead of assuming a well defined flux, for shallow faint end source counts such as here.  

{\it Radio incompleteness}: Lastly, the selection function for the FIRST objects in the White et al. sample used here might not be a sharp Heavyside function at a peak pixek flux density of 1 mJy, but rather smeared out.  According to Figure 1 of \citet{Jiang07}, the selection function of FIRST for SDSS optically identified quasars is such that at an integrated flux density of 1 mJy only about 55\% of sources are seen, and this number rises to 75\% at 1.5 mJy and about 85\% at 2 mJy.\footnote{The fuzzyness of the truncation boundary has a similar effect as the data measurement errors in the sense that it is unimportant for $m_{below}=-1$ and more important for larger deviations from this value.}  This particular selection function would likely not be identical for the POSS-I optically identified quasars of the sample we use here.  Also, we have considered the sample to be limited by the peak pixel flux (i.e. surface brightness limited) in the radio rather than being limited by the integrated flux, in accordance with the criteria set forward in White et al.  So it is difficult to directly compare the potential radio selection function here with the one in Jiang et al.  

The way to test the effects on our analysis is to repeat the analysis limiting the sample to a higher radio flux, where the sample would presumably be more complete, and determine the extent to which the calculated parameters change in a systematic way.  We have done so with lower radio flux limit of 2 mJy (486 objects), as opposed to the original 1 mJy (636 objects).  The effect. propagated through the analysis, is primarily to extend the 1$\sigma$ uncertainties on $k_{\rm opt}$  and $k_{\rm crr}$ in the direction of lower $k_{\rm opt}$ (1.25 on the low end) and higher $k_{\rm crr}$ (3.75 on the high end), and to move the best bit values to 2 and 2.5 respectively.  There is no discernable effect on the value of the correlation parameter $\alpha$.  The modified best fit values and increased 1$\sigma$ uncertainty for $k_{\rm opt}$ and $k_{\rm crr}$ only slightly alters the 1$\sigma$ range and the best fit value of $k_{\rm rad}$, since the particular error ellipse shape means that lower $k_{\rm opt}$ values accompany higher $k_{\rm crr}$ values.  The main effect on physical parameters then is to shift the best fit and low end 1$\sigma$ values for $k_{\rm opt}$ downward. We also find that there is a negligible effect on the density evolutions, there is a small flattening effect on the high end optical and radio LF slopes, which are reported in \S \ref{local}, and a negligible effect on the shape of the $G_{R}\!(R)$ distribution.  To the extent that faint flux radio incompleteness is present in the sample considered here, it does not seem to have a large systematic effect on the determination of the parameters in this analysis.

\section{Discussion}

We have used a general and robust method to determine the radio and optical luminosity evolutions simultaneously for the quasars in the \citet{White00} dataset, which combines 1.4 GHz radio and $R$-band optical data for 636 quasars ranging in redshifts from 0.02 to 3.425 and over seven orders of magnitude in radio loudness.  We find that the quasars exhibit more substantial radio evolution than optical evolution with redshift (\S \ref{method} and Figure \ref{alphas}).  We also show that when divided into RL and RQ sets accordingly to the standard definition (divided by the value of the radio-loudness parameter $R=10$),  the two sub-populations exhibit similar density evolution.  The local optical and radio LFs that we obtain are consistent with previous determinations.   

Differences are noted with previous determinations of the radio luminosity evolution of quasars.  \citet{Willott01} also use a power law parameterization of the radio evolution with redshift for the radio bright sources they consider.  Our result for the radio luminosity evolution, when evaluated for the data set as a whole is not consistent to within uncertainty with their results (power laws ranging from 3.1 to 3.6).  \citet{Strazz10} have recently obtained results with a radio survey to low (13.5 $\mu$Jy) flux limits, quoting $k_{\rm rad}$ = 2.7 $\pm$ 0.3 with the same parameterization used here for the sub-population that they identify as AGN.  This is also not in agreement with the radio evolution determined in our analysis.   In a future paper we intend to carry out the same analysis using the much larger SDSS sample of quasars used by \citet{Jiang07}.

There has been much discussion as to whether RL and RQ quasars, defined solely by means of the radio loudness parameter as explained above, constitute a true continuum or two populations that can be said to be distinct in some way (e.g. \citet{Kellerman89}; \citet{Ivezic02}; \citet{Cira03}; \citet{Ivezic04}).  Our analysis favors the former.  First, we found that the division of the quasar population into the two aforementioned classes introduces strong biases into the simultaneous determination of the radio and optical luminosity evolutions (\S \ref{method}; see also Figure \ref{alphas}).   More importantly, as shown in \S \ref{Rdist}, forming a distribution in the raw values of the radio loudness parameter $R$, without taking into account the biases introduced by the truncations in the data and the correlated luminosity evolutions, results in a shape for the distribution which is very different from the true distribution; e.g. it shows a possible dip in $G_R(R)$ at $R \sim 10$, while the true distribution is rather smooth and shows at most a modest feature.  Even if the feature is real, it does not suggest any bi-modality, but rather a continuous range of physical properties in a single population. We also find that at higher redshifts and optical luminosities the radio loudness become more pronounced. This is opposite of the trends presented by \citet{Jiang07}, but in agreement with others such as \citet{Donoso09}.

Accessing the unbiased distribution of the radio loudness parameter for quasar sources is crucial not only for understanding the cosmological evolution of this class of active galaxies, but also for understanding jet launching processes in the vicinities of supermassive black holes. In this context, we note that the observed optical fluxes of quasars are dominated by the emission from an accretion disk accreting at relatively high rates, around $1\%-100\%$ of Eddington, and therefore radiating with $\simeq 10\%$ efficiency.  Hence, the optical luminosity is a very good measure of the total accretion power in quasar sources.  On the other hand, the observed radio fluxes of the discussed class of objects are expected to originate in the outflowing magnetized plasma. In particular, the radio emission of quasars is produced predominantly via the synchrotron emission of relativistic well-collimated jets (in the case of very radio loud sources), or via the cyclotron and/or free-free emission of at most mildly-relativistic disk winds (in the case of very radio quiet nuclei). In both cases, the observed radio luminosities should be considered as proxies for the kinetic luminosities of the outflowing matter. Therefore, the radio loudness $R$ characterizes the efficiency of the production of jets/outflows for a given accretion power.  

The lack of any clear bi-modality centered around $R$=10 in the distribution of the radio loudness parameter for quasars, as advocated here, implies then that there is no critical change in the parameters of the central engine between the RL quasars (those producing extremely powerful relativistic jets), and the RQ ones (those producing only mildly-relativistic and uncollimated disk winds). This is a crucial piece of evidence for understanding still debated mechanisms for jet launching in AGNs. Note, for example, that our finding is hardly consistent with the idea that RL quasars possess counter-rotating (with respect to the black hole spin) accretion disks, as opposed to RQ quasars with co-rotating disks only.  Instead, the nuclei of jetted and non-jetted quasar sources seem to be intrinsically very similar, differing only smoothly and continuously in some particular respects. But it has to be emphasized that here we do not discuss the whole population of AGNs (including, e.g., Seyfert galaxies) but only strictly the quasar population \citep[see the related recent discussion in][and references therein]{Sikora07}.  We note that our analysis does not address the question of whether there may be a large population of quasars with values of $R$ beyond the range of the present sample, in particular we cannot rule out a significant population with very low values of $R$ ($\log R <$ 0.01). 

Another (possibly related) result we find is that the radio loudness may increase with increasing optical and radio luminosities, as the best fit value for the correlation parameter $\alpha$ is 1.3, although a strictly linear correlation is not ruled out with much significance, and furthermore, one may dispute that the correlation is inherent in the population (see the discussion in Appendix B).  However, if this super-linear correlation were indeed intrinsically the case, it would imply the existence of some connection between the efficiency of formation of relativistic jets and accretion power, which may in turn depend on the combination of the evolving accretion rate and black hole spin \citep[see in this context][]{Tschaikovsky09}, given the observed increase in radio loudness with redshift.  Note, however, that even though we have used simple one-parameter functions to describe the emerging correlations, it is possible that some of them are more complex.  For example, the correlation index $\alpha$ between the radio and optical luminosities may be close to unity only for low luminosity objects, but much larger than that for more luminous (and therefore radio loud) quasars. More data and further analysis is needed to address this and similar issues, which may provide further constraints on theoretical models.

Another application of the presented analysis is related to the understanding of the origin of the cosmic background radiation in the radio frequency regime.  In \citet{Singal10} we estimated the fractional contribution of quasars to the Cosmic Radio Background, assuming the level reported by \citet{Fixsen10}. In general the flux of the objects fitting the definition of radio loud is well characterized by current interferometric radio surveys so that their contribution to the total radio background intensity can be estimated to be 15\% to 25\% of the observed value.  In the earlier work we also estimated the total contribution to the background of the radio quiet objects, and found it to be between 1\% and 2\% for favored models of quasar luminosity evolution.  This estimate was based on integrating values of the quasar bolometric LF, as reported in the literature, over redshift, applying a mapping between optical and radio luminosity, and assuming that the optical and radio luminosities had identical redshift evolutions.  We also noted there that the contribution we estimated was dependent on the later assumption and would be revised in the case of differing radio and optical luminosity evolutions.  As we see here that the population of quasars has greater radio evolution relative to optical, the contribution of RQ quasars to the radio background will be somewhat larger than the value reported in our previous work.  We will present a quantitative determination in a forthcoming paper.

\acknowledgments

JS thanks S. Kahn and R. Schindler for their encouragement and support.  {\L}S was supported by the Polish Ministry of Science and Higher Education through the project N N203 380336.

\clearpage

\appendix

\section*{Appendix A}

{\bf Distribution of Radio Loudness $R$ }: As described in the introduction one of the aims of this paper is determination of the distribution of the radio to optical luminosity ratio $R= L_{\rm rad}/L_{\rm opt}$ from an observed sample of radio and optical fluxes and redshifts. This requires a proper accounting for correlations and evolutions of optical and radio luminosities and the observational selection effects.  Here we describe the how these factors affect the observed distribution of $R$.

The true or intrinsic distribution of $R$ values is related to the radio and optical LFs
$\Psi (L_{\rm opt},  L_{\rm rad}, z)$ as
\begin{equation}
G_T(R,z)=\int^\infty_0\Psi (L_{\rm opt},  R \,\, L_{\rm opt}, z) \, L_{\rm opt} \,
dL_{\rm opt}=\int^\infty_0\Psi \left( {L_{\rm rad} \over R},  L_{\rm rad}, z \right) \,
L_{\rm rad} \, {{
dL_{\rm rad} } \over {R^2}}.
\label{Gtrue2}
\end{equation}
(compare to the separable, local form in Equation \ref{localr})\\

The observed distribution on the other hand is different because the observational selection effects truncate the data. For example for a sample with well defined flux limits  $f_{\rm rad}\geq f_{m,rad},\, f_{\rm opt}\geq f_{m,opt}$, and $R_{\rm obs}\equiv f_{m,rad}/f_{m,opt}$ the observed distributions is 
\begin{eqnarray}
G_{\rm obs}(R,z) & = & \int^\infty_{L_{\rm min,opt}(z)}\Psi (L_{\rm opt},  R \,\,
L_{\rm opt}, z) \, L_{\rm opt} \, dL_{\rm opt}\,\,\,\,\, {\rm for}\,\,\,\,\, R>R_{\rm obs}\\
G_{\rm obs}(R,z )& = & \int^\infty_{L_{\rm min,rad}(z)}\Psi \left({L_{\rm rad} \over
R}, 
L_{\rm rad}, z \right) \, L_{\rm rad} \, {{ dL_{\rm rad} } \over {R^2}} \nonumber \\
& = & \int^\infty_{L_{\rm min,opt}(z)R_{\rm obs}/{\hat R}}\Psi (L_{\rm opt},  R \,\,
L_{\rm opt}, z) \, L_{\rm opt} \, dL_{\rm opt}\,\,\,\,\, {\rm for}\,\,\,\,\,
R<R_{\rm obs},
\label{GandPsi}
\end{eqnarray}
where 
\begin{equation}
{\hat R}=(K_{\rm opt}/K_{\rm rad})R,\,\,\,\,L_{\rm min,opt}(z)=4\pi d_L^2(z)K_{\rm opt}
 f_{m,opt}\,\,\,\,\,
{\rm and}\,\,\,\,\,L_{\rm min,rad}(z)=4\pi d_L^2(z)K_{\rm rad} f_{m,rad}.
\label{lminz}
\end{equation}
and $K_i$ is the K-correction factor for waveband $i$.  All of these are obtained from the observed distributions of the fluxes and redshifts. If we approximate the observed distribution of fluxes and redshifts by a continuous function
$n_{\rm obs}(f_{\rm opt},f_{\rm rad},z)$ then
\begin{equation}
\Psi (L_{\rm opt},  L_{\rm rad}, z)=n_{\rm obs}\left({L_{\rm opt}\over 4\pi
d_L^2(z)K_{\rm opt}},{L_{\rm rad}\over
4\pi d_L^2(z)K_{\rm rad}},z\right )\left({1\over 4\pi d_L^2}\right )^{2} {1\over
K_{\rm opt}K_{\rm rad}V'(z)}
\label{psiandn}
\end{equation}
so that 
\begin{eqnarray}
G_T(R,z)  & = & (K_{\rm opt}K_{\rm rad}V')^{-1}\int^\infty_0 n_{\rm obs}(f_{\rm opt},{\hat
R}f_{\rm opt},z) \, f_{\rm opt} \, df_{\rm opt} \\
G_{\rm obs}(R,z) & = & (K_{\rm opt}K_{\rm rad}V')^{-1}\int^\infty_{f_{lim}} n_{\rm
obs}(f_{\rm opt},{\hat R}f_{\rm opt},z) \, f_{\rm opt} \, df_{\rm opt},
\label{Gandn}
\end{eqnarray}
where  $V'=dV(z)/dz$,   and $f_{lim}=f_{m,opt}$ for ${\hat R}>R_{\rm obs}$ and $f_{lim}=f_{m,opt}(R_{\rm obs}/{\hat R})$ for ${\hat R}<R_{\rm obs}$. Note that ${\hat R}$ depends on redshift to the extent that the optical and radio K-corrections are different. However, since the radio and optical spectra can be well approximated by power laws with the almost same spectral index the two K-corrections are almost equal. In what follows we ignore this small difference and set ${\hat R}=R$. The sources with $R>R_{\rm obs}$ can be called {\it optically limited} because their optical flux is the main determining factor for their inclusion in the sample. Similarly sources with $R<R_{\rm obs}$ can be called {\it radio limited}.

Clearly the intrinsic and observed distributions of $R$ are different.  In reality the situation is slightly more complicated because the lower limit of the integration does not extend to zero. The samples of available quasars are truncated also by  minimum luminosities, say $L_{m,opt}$ and $L_{m,rad}$ which introduces a second critical value for $R$, namely $R_{\rm int}\equiv L_{m,rad}/L_{m, opt}$. The above equations are valid for redsifts $z>z_{\rm min,opt}$ or $z_{\rm min,rad}$ defined as 
\begin{equation}
L_{m,i}=4\pi d_L^2(z_{\rm min,i})K_i(z_{\rm min,i})f_{m,i}.
\label{zmin}
\end{equation}
For $z<z_{min}$ (defined as the lower of $z_{\rm min,opt}$ and $z_{\rm min,rad}$) there is no truncation due to flux limits and $G_{\rm obs}=G_T$.\\

It is  convenient to define
\begin{equation}
\Phi(R,z;x)=\int^\infty_i\Psi(L_{\rm opt},R \,\, L_{\rm opt},z) \, L_{\rm opt} \, dL_{\rm opt}
\label{Phi}
\end{equation}
so that the true distribution can be written as
\begin{eqnarray}
G_T(R,z) & = & \Phi(R,z;L_{m,opt})  \,\,\,\,\,{\rm  for}\,\,\,\,\, R>R_{\rm
int}\\
G_T(R,z) & = & \Phi \left(R,z; {{R_{\rm int} \, L_{m,opt}} \over {R}} \right)
\,\,\,\,\,{\rm
for}\,\,\,\,\,R<R_{\rm int}.
\end{eqnarray}
The observed distribution depends on the relative values of $R_{\rm int}$ and
$R_{\rm obs}$.\\

For a sample where $R_{\rm obs}>R_{\rm int}$, which means that $z_{\rm min,rad}>z_{\rm min,opt}$, such a sample may be classified as {\it mainly optically selected}. (For purely optically selected sample  $R_{\rm obs}=0$.) In this case observed distribution will be the same as the true distribution ($G_{\rm obs}=G_T$) for all redshifts $z<z_{\rm min,rad}$.  Obviously, the reverse is true in the opposite case with radio exchanged for optical. At higher redsifts
\begin{eqnarray}
{G_{\rm obs}(R,z)\over G_T} & = & {{ \Phi(R ,z ;L_{\rm min,opt}(z))} \over
{\Phi(R, z; L_{m,opt}) }}\,\,\,\,\,{\rm  for}\,\,\,\,\, R>R_{\rm
int}\\
{G_{\rm obs}(R,z)\over G_T} & = &  {\Phi \left(R ,z ;{R_{\rm obs}\,L_{\rm
min,opt}(z)\over R} \right) \over \Phi \left(R, z; {R_{\rm int} \, L_{m,opt}
\over {R}} \right) } \,\,\,\,\,{\rm
for}\,\,\,\,\,R<R_{\rm int}.
\end{eqnarray}\\

{\bf A simple Example}: Let us assume that the radio and optical luminosities are uncorrelated and do not evolve so that we have $\Psi (L_{\rm opt},  L_{\rm rad}, z)=\psi_{\rm opt}(L_{\rm opt}) \, \psi_{\rm rad}(L_{\rm rad}) \, \rho(z)$, where $\rho(z)$ describes the density evolution.  Furthermore if we assume simple power law LFs $\psi_{\rm opt}(L_{\rm opt})=A_{\rm opt}L_{\rm opt}^{-\alpha_{\rm opt}}$ and $\psi_{\rm rad}(L_{\rm rad})=A_{\rm rad}L_{\rm rad}^{-\alpha_{\rm rad}}$, it is easy to show that 
\begin{equation}
G_{T}(R,z)\propto  (R/R_{\rm int})^{1-\alpha_{\rm rad}} \,\,\,\,{\rm
for}\,\,\,\,
R>R_{\rm int}\,\,\,{\rm and}\,\,\,\,\, G_{T}(R,z)\propto  (R/R_{\rm
int})^{\alpha_{\rm opt}-1}
\,\,\,\,{\rm for}\,\,\,\, R<R_{\rm int}.
\label{Gtrue3}
\end{equation}
Similarly, it is easy to show that for redshifts $z>z_{min}$

$$G_{\rm obs}(R,z)=G_T{\left(L_{\rm min,opt}(z) \over  L_{m,opt}\right)^\beta}
\times
 \cases {1  &for 
$R>R_{\rm obs}>R_{\rm int}$, \cr
 (R_{\rm obs}/R)^{\beta} &for
$R_{\rm obs}<R<R_{\rm int},$\cr
 (R_{\rm obs}/R_{\rm int})^\beta &for
 $R_{\rm obs}<R_{\rm int}<R$\cr
}$$\\

where $\beta=\alpha_{\rm opt}+\alpha_{\rm rad}-2$.

The observed and true distributions have different shapes in the range between the intrinsically and observationally limiting values of $R$. The shapes become identical when these two values are equal. Larger differences will be the case if the radio and optical luminosities are correlated non-linearly and undergo different kinds of luminosity evolution.  As a result the fraction of RL or RQ sources (arbitrarily chosen at some value or $R$) will vary with redshift and/or luminosities.  For example if the radio and optical luminosities were correlated linearly making the the radio loudness independent of both luminosities then we will be dealing with a separable LF $\psi(R, L_{\rm opt},z )=G(R) \psi_{\rm opt}(L_{\rm opt},z)$. In such a case, because of flux limits at any redshift the observed range of extends to $R_{\rm min}(L_{\rm opt},z )$ which decreases with increasing $L_{\rm opt}$ but increases with increasing $z$. This will cause the fraction of RL sources to decrease with luminosity but increase with redshift.  All such trends can be determined by proper accounting of the correlations and evolution as described in this paper.

\section*{Appendix B}

\begin{figure}
\includegraphics[width=3.5in]{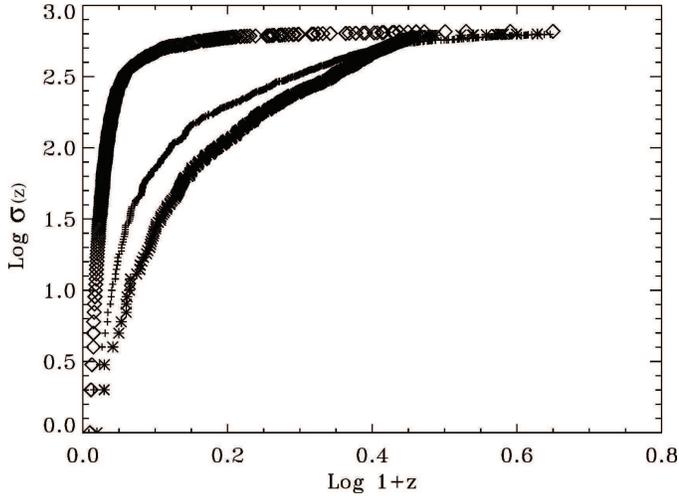}
\caption{Raw observed cumulative distribution in redshift of the White et al. dataset (small plusses), the `observed' dataset from the uncorrelated MC population (diamonds), and the `observed' dataset from the correlated MC population (stars).  The normalization is arbitrary.  The redshift distribution of the observed correlated MC population much more closely resembles that of the real observed data. }
\label{zdizists}
\end{figure}

{\bf Luminosity correlations and flux limits:} As pointed out by the referee, it has been suggested by \citet{Ski} that the observed correlation between the luminosities may be induced by observational selection effects.  Other works that have examined the issue include \citet{Khembavi86}, \citet{FB83} and \citet{Chanan83}.  It is not clear whether the correlation seen in this work between $L_{\rm rad}$ and $L_{\rm opt}$ is inherent in the quasar population, or is introduced by the selection effects of the surveys.  In some sense this question does not matter for the analysis presented here, because the rotation to $L_{\rm crr}$ is a technique required to achieve independent variables ($L_{\rm opt}$ and $L_{\rm crr}$) in the context of the data present so that we can recover the inherent redshift evolutions.  This is independent of the underlying luminosity-luminosity correlation of the population.  However, it is of general interest whether the correlation seen between radio and optical luminosity is inherent.

In order to begin an investigation of this, we simulate via Monte Carlo (MC) techniques two cases of quasar populations, distributing the objects in radio luminosity, optical luminosity, and redshift, and applying the flux limits of the White et al. dataset to achieve for both MC populations `observed' datasets of a similar size to the real White et al. data used in this analysis.  For the MC populations we use a redshift distribution with $\rho\!(z) \propto z^{4}$ to redshift 1.2 and then constant density with redshift for $1.2 \leq z \leq 3.2$.  In the first, uncorrelated, case, we draw the optical and radio luminosities of the population from separate distributions with power law slopes in luminosity of $-$2.  In the second, correlated, case, we draw the optical luminosities of the population from a distribution with a power law slope in luminosity of $-$2, then the radio luminosities of the population are assigned according to $L_{\rm rad} \propto {L_{\rm opt}}^{1.3}$ and then randomized about that value by a factor of $10^{2.5 \times A}$ where A is a normally distributed random number with mean of zero and standard deviation of one.  For this simple analysis we have not included the effects of luminosity evolution, because it would require an orders of magnitude larger simulation.

Our results show that that indeed the radio and optical luminosities of the 'observed' datasets in both cases are correlated.  An analysis identical to that presented in \S \ref{roevsec} reveals that the observed data for the correlated MC population has a value for the correlation index $\alpha$ of $1.3 \pm 0.3$ while that of the uncorrelated MC population has a value of $0.85 \pm 0.15$.  However, these `observed' sets can be compared to the real White et. al dataset, and it is seen that the observed dataset for the correlated MC population more closely resembles the real dataset than that of the uncorrelated MC population does, for instance comparing the redshift distributions of the observed objects, shown in Figure \ref{zdizists}.  From this analysis we conclude that the correlation seen in the White et al. dataset between $L_{\rm rad}$ and $L_{\rm opt}$ could be inherent in the population.  Exact determinations of what fraction of the observed correlations is inherent in the population and what fraction is due to selection effects depends on the values of the many parameters describing the LFs and evolutions.  This is an important issue in many areas of astrophysics and requires considerable work which is beyond the scope of this paper.  In a forthcoming work we will address this general question.


\begin{thebibliography}{}

\bibitem[Antonucci(2011)]{Ski} Antonucci, R. 2011 (arXiv:1101.0837)
\bibitem[Boyle et al.(2000)]{Boyle00} Boyle, B., Shanks, S., Croom, R., Smith, L., Loaring, N. \& Heymans, C. 2000, \mnras, 317, 1014
\bibitem[Caditz \& Petrosian(1993)]{CP93} Caditz, J. \& Petrosian, V. 1993, \apj, 416, 450
\bibitem[Chanan (1983)]{Chanan83} Chanan, G. 1983, \apj, 275, 45
\bibitem[Cirasuolo et al.(2003)]{Cira03} Cirasuolo, M., Magliocchetti, A. \& Danese, L. 2003, \mnras, 341, 993
\bibitem[Cirasuolo et al.(2006)]{Cira06} Cirasuolo, M., Magliocchetti, A., Gentile, G., Celotti, A., Christiani, S. \& Danese, L. 2006, \mnras, 371, 675
\bibitem[Croom et al.(2009)]{Croom09} Croom, S., et al. 2009, \mnras, 399, 1755
\bibitem[Donoso et al.(2009)]{Donoso09} Donoso, E., Best, P., \& Kauffmann, G. 2009, \mnras, 392, 617
\bibitem[Dunlop \& Peacock(1990)]{DP90} Dunlop, J. \& Peacock, J. 1990, \mnras, 247, 19
\bibitem[Efron \& Petrosian(1992)]{EP92} Efron, B. \& Petrosian, V. 1992, \apj, 399, 345
\bibitem[Efron \& Petrosian(1999)]{EP99} Efron, B. \& Petrosian, V. 1999, JASA, 94, 447
\bibitem[Feigelson \& Berg(1983)]{FB83} Feigelson, E. \& Berg, C. 1983, \apj, 269, 400
\bibitem[Fixsen et al.(2010)]{Fixsen10} Fixsen D. et al., 2010, \apj, 2011, 734, 5
\bibitem[Goldschmidt et al.(1999)]{Goldschmidt99} Goldschmidt, P., Kukula, M., Miller, L., \& Dunlop, J. 1999, \apj, 511, 612
\bibitem[Hewett et al.(2001)]{Hewett01} Hewett, P., Foltz, C. \& Chaffe, F. 2001, \apj, 122, 518
\bibitem[Hopkins et al.(2007)]{Hopkins07} Hopkins, P., Richards, G. \& Hernquist, L. 2007, \aj, 654, 731
\bibitem[Hopkins et al.(2010)]{Hopkins10} Hopkins, P., Younger, J., Hayward, C., Narayan, D., \& Hernquist, L. 2010, \mnras, 402, 1693
\bibitem[Ivezic et al.(2002)]{Ivezic02} Ivezic, Z. et al. 2002, \aj, 124, 2364
\bibitem[Ivezic et al.(2004)]{Ivezic04} Ivezic, Z. et al. 2004, in AGN Physics With the SDSS, eds. G.T. Richards \& P.B. Hall (San Francisco: ASP), 347
\bibitem[Jiang et al.(2007)]{Jiang07} Jiang, L., et al. 2007, \apj, 656, 680
\bibitem[Kellerman et al.(1989)]{Kellerman89} Kellerman, K., Sramek, R., Schmidt, M., Shaffer, D. \& Green, R. 1989, \aj, 98, 1195
\bibitem[Khembavi et al.(1986)]{Khembavi86} Khembavi, A., Feigelson, E., \& Singh, K. 1986, \mnras, 220, 51
\bibitem[LaFranca et al.(2010)]{LaFranca10} LaFranca, F., Melini, G., \& Fiore, F. 2010, \apj, 718, 368
\bibitem[Lynden-Bell(1971)]{L-B71} Lynden-Bell, B. 1971, \mnras, 155, 95
\bibitem[Maloney \& Petrosian(1999)]{MP99} Maloney, A. \& Petrosian, V. 1999, \apj, 518, 32
\bibitem[Marshall et al.(1983)]{Marshall83} Marshall, H., Tananbaum, H., Avni, Y., \& Zamorani, G. 1983, \apj, 269, 35
\bibitem[Mauch \& Sadler(2007)]{MS07} Mauch, T. \& Sadler, E. 2007, \mnras, 375, 931
\bibitem[Matute et al.(2006)]{Matute06} Matute, I., LaFranca, F., Pozzi, F., Gruppioni, C., Lari, C \& Zamorani, G. 2006, \aap, 451, 443
\bibitem[Miller et al.(1990)]{Miller90} Miller, L., Peacock, J., \& Mead, A. 1990, \mnras, 244, 207
\bibitem[Osterbrock (1989)]{Osterbrock89} Osterbrock, D., {\it `Astrophysics of Gaseous Gebulae and Active Galactic Nuclei
'}, Mill Valley, CA: University Science Books 1989
\bibitem[Petrosian(1992)]{P92} Petrosian, V. 1992, in Statistical Challenges in Modern Astronomy, ed. E.D. Feigelson \& G.H. Babu (New York:Springer), 173
\bibitem[Petrosian(1973)]{P73} Petrosian, V. 1973, \apj, 183, 359
\bibitem[Richards et al.(2006)]{Richards06} Richards, G. et al. 2006, \aj, 131, 2766
\bibitem[Schmidt (1972)]{Schmidt72} Schmidt, M. 1972, \apjs, 176, 273
\bibitem[Schmidt (1967)]{Schmidt68} Schmidt, M. 1968, \apj, 151, 393
\bibitem[Shaver et al.(1996)]{Shaver96} Shaver, P., Wall, J., Kellermann, K., Jackson, C., \& Hawkins, M. 1996, \nat, 384, 439
\bibitem[Singal et al.(2010)]{Singal10} Singal, J., Stawarz, {\L}., Lawrence, A., \& Petrosian, V. 2010, \mnras, 409, 1172
\bibitem[Sikora et al.(2007)]{Sikora07} Sikora, M., Stawarz, {\L}., \& Lasota, J.-P. 2007, \apj, 658, 815
\bibitem[Strazzullo et al.(2010)]{Strazz10} Strazzullo, V., Pannella, M., Owen, F., Bender, R., Morrison, G, Wei-Hao, W. \& Shupe, D. 2010, \apj, 714, 1305
\bibitem[Tchekhovskoy et al.(2010)]{Tschaikovsky09} Tchekhovskoy, A.,Narayan, R., \& McKinney, J.  2010, \apj, 711, 50
\bibitem[Ueda et al.(2003)]{Ueda03} Ueda, Y., Akiyama, M., Ohta, K., \& Miyaji, T. 2003, \apj, 598, 886
\bibitem[White et al.(2000)]{White00} White, R. et al. 2000, \apjs, 126, 133
\bibitem[Willott et al.(2001)]{Willott01} Willottt, C., Rawlings, S., Blundell, K., Lacy, M., \& Eales, S. 2001, \mnras, 322, 536

\end{thebibliography}
\end{document}